\documentstyle[12pt,aaspp4]{article}

\begin{document}

\newcommand{\gsim}{\mbox{\raisebox{-1.0ex}{$~\stackrel{\textstyle >}
{\textstyle \sim}~$ }}}
\newcommand{\lsim}{\mbox{\raisebox{-1.0ex}{$~\stackrel{\textstyle <}
{\textstyle \sim}~$ }}}
\newcommand{\psim}{\mbox{\raisebox{-1.0ex}{$~\stackrel{\textstyle \propto}
{\textstyle \sim}~$ }}}
\newcommand{\vect}[1]{\mbox{\boldmath${#1}$}}
\newcommand{\lmk}{\left(}
\newcommand{\rmk}{\right)}
\newcommand{\lnk}{\left\{ }
\newcommand{\nn}{\nonumber}
\newcommand{\rnk}{\right\} }
\newcommand{\lkk}{\left[}
\newcommand{\rkk}{\right]}
\newcommand{\lla}{\left\langle}
\newcommand{\p}{\partial}
\newcommand{\rra}{\right\rangle}
\newcommand{\vex}{{\vect x}}
\newcommand{\vek}{{\vect k}}
\newcommand{\vel}{{\vect l}}
\newcommand{\vem}{{\vect m}}
\newcommand{\ven}{{\vect n}}
\newcommand{\vep}{{\vect p}}
\newcommand{\veq}{{\vect q}}
\newcommand{\veX}{{\vect X}}
\newcommand{\veV}{{\vect V}}
\newcommand{\beq}{\begin{equation}}
\newcommand{\eeq}{\end{equation}}
\newcommand{\beqa}{\begin{eqnarray}}
\newcommand{\eeqa}{\end{eqnarray}}
\newcommand{\mpc}{\rm Mpc}
\newcommand{\hmpc}{{h^{-1}\rm Mpc}}
\newcommand{\ch}{{\cal H}}

\title{ Nonlinear Velocity-Density Coupling: \\Analysis by Second-Order
  Perturbation Theory  }
\author{\sc Naoki Seto }
\affil{Department of Physics, Faculty of Science, Kyoto University,
Kyoto 606-8502, Japan\\
seto@tap.scphys.kyoto-u.ac.jp }

\begin{abstract}
Cosmological linear perturbation theory predicts that the peculiar velocity
$\veV(\vex)$ and the matter overdensity $\delta(\vex)$ at a same point $\vex$
are statistically independent quantities, as log as the initial density
fluctuations are random Gaussian distributed. However nonlinear
gravitational effects might change the situation. Using  framework of
second-order perturbation theory and the Edgeworth expansion method, we
study local density dependence of bulk velocity dispersion that is
 coarse-grained at a weakly nonlinear scale. For a typical CDM model, the first
nonlinear correction of this constrained bulk velocity dispersion
amounts to  $\sim 0.3\delta$ (Gaussian smoothing)  at a  weakly nonlinear scale
 with a very  weak dependence on cosmological
parameters.     We also compare our analytical prediction with published 
numerical results given at nonlinear regimes.

\end{abstract}

\section{Introduction}
The peculiar velocity field is one of the most fundamental quantities to
analyze the 
large-scale structure in the universe ({\it e.g.} Peebles 1980). It is
considered to  reflect 
dynamical nature of density fluctuations of gravitational
matter. The peculiar velocity field is usually observed  using
  astrophysical objects ({\it e.g.} galaxies), as  determination of
distances is  crucial for measuring  peculiar velocities (Dekel 1994,
Strauss \& Willick 1995). There is a
possibility that statistical aspects of the  velocity field traced by these
objects and that traced  by dark matter particles might be different.  This
difference is generally called ``velocity bias'' and its elucidation becomes
highly important in observational cosmology ({\it e.g.} Cen \& Ostriker 1992,
Narayanan, Berlin \& Weinberg 1998, Kaufmann et al. 1999).

Velocity bias is often discussed numerically with making ``galaxy 
particles" in some effective manners. But here we discussed  a more basic
phenomenon. It is known that  
statistics of the peculiar velocity field depend largely on local density
contrast. For example, both the single particle and
pairwise velocity dispersions   of dark matter particles are 
known to be increasing function of local density (Kepner, Summers, \&
Strauss 1997, Strauss, Cen \& Ostriker 1998, Narayanan et al. 1998). 
Analysis of pairwise velocity statistics  is  interesting from theoretical 
point of views, and also very important in observational cosmology
(Peebles 1976, Davis \& Peebles 1983, Zurek et al. 1994, Fisher et
al. 1994,  Sheth 1996,
Diaferio \& 
Geller 1996, Suto \& Jing 1997, Seto \& Yokoyama 1998a, 1999, Jing \&
B$\ddot{\rm o}$rner 1998, 
Juszkiewicz, Fisher \& 
Szapudi 1998, Seto 1999b,  Juszkiewicz, Springel \&
Durrer 1999). But 
we do not discuss it here and concentrate on velocity field 
characterized by single point  which is simpler to analyze
theoretically.  

Linear perturbation theory predicts that the peculiar velocity
$\veV(\vex)$ and the density contrast $\delta(\vex)$ at a given point
$\vex$ is statistically independent,  as long as the  initial density
fluctuations are  random Gaussian distributed. Namely, the joint probability
distribution function $P(\veV, \delta)$ can be written in a  form 
as $P_1(\veV)P_2(\delta)$.

It is not surprising that the  peculiar velocity of each particle is largely
affected by nonlinear gravitational effects and shows local density
dependence described above.  But what can we expect for the smoothed (bulk)
velocity that is 
field coarse grained at  some spatial scale $R$?  Due to nonlinear mode
couplings, the relation   $P_1(\veV)P_2(\delta)$ valid for linear theory must
be modified and  bulk velocity dispersion must also  depend on local
density contrast defined at the same  smoothing scale $R$ (see
Bernardeau 1992,
Chodorowski \& {\L}okas 1997, Bernardeau et al. 1999 for the velocity
divergence field). 

However, Kepner, Summers \& Strauss (1997) showed from cold-dark-matter
 (CDM) and hot-dark-matter (HDM)
 N-body simulations  that at nonlinear scales  ($0.77\hmpc\le R \le
 4.88\hmpc$), such a local density dependence was not observed (see
 Figs.2(a) and 3(a) of their paper). 
This is an interesting contrast to the behavior of velocity field traced 
 by each particle, as described before (Kepner et al. 1997, Narayanan et 
 al. 1998).

In this article, we investigate local density dependence of smoothed
(bulk) velocity dispersion using framework of second-order perturbation
theory. We calculate the first-order nonlinear correction of the constrained
 velocity
dispersion.  Our target is weakly nonlinear scale and somewhat larger
than scale analyzed by Kepner et al. (1997). Since current  survey depth
of the
cosmic  velocity field is highly limited, our constrained statistics
might not be useful in observational cosmology at present ({\it e.g.}
Seto \& Yokoyama 1998b, see also Seto 1999a). Our interest in this
article  is   
theoretically motivated one about  nonlinear gravitational dynamics.

As  the peculiar velocity field is more weighted to large-scale fluctuations
(smaller wave number $k$) than the density field, perturbative treatment of
smoothed velocity field 
would  be reasonable at weakly nonlinear scale.  Actually, Bahcall,
Gramann \& Cen (1994) showed that 
smoothed unconstrained 
velocity dispersion  in N-body simulations are well predicted by 
linear theory even at smoothing scale
$R=3\hmpc$ (see their Table 1).  Second-order
analysis by Makino, Sasaki \&
Suto (1992) also gives consistent results to their simulations.

\section{Formulation}
First we define  the (unsmoothed) density contrast field $\delta(\vex)$
in terms 
of the mean density of the universe $\bar{\rho}$ and the local density
field $\rho(\vex)$  as
\beq
\delta(\vex)=\frac{\rho(\vex)-\bar{\rho}}{\bar{\rho}}.
\eeq
Many theoretical predictions of the large-scale structure are based on
continuous fields, but observations as well as numerical experiments
(such as,  N-body simulations)  are usually sampled by  points 
where  point-like
galaxies (or mass elements) exist.  In comparison  of theoretical
predictions with actual
observations or numerical experiments,  
smoothing operation  becomes  sometimes 
crucially important to remove sparseness of particles' system.  This
operation is also important to reduce strong nonlinear effects which are 
difficult to handle theoretically.  Thus it is favorable to make theoretical
predictions of the large-scale structure including smoothing operation.
We can express the smoothed density
contrast field  $\delta_R(\vex)$  and the smoothed velocity field 
$\veV_R(\vex)$ with (spatially
isotropic) filter $W(x,R)$ as   
\beq
\delta_R(\vex)\equiv \int d\vex'^3 \delta(\vex') W(|\vex-\vex'|,R),~~~
\veV_R(\vex)\equiv \int d\vex'^3 \veV(\vex') W(|\vex-\vex'|,R).
\eeq
As we discuss only the  smoothed  fields in this article, we hereafter 
omit  the suffix $R$ which indicates smoothing radius. 

The velocity dispersion $ \Sigma_V^2 (\delta)$ for points $\vex$ with a
given 
overdensity $\delta(\vex)=\delta$ is formally written as 
\beq
 \Sigma_V^2 (\delta)=\frac{\lla \veV(\vex)^2\delta_{Drc}[\delta(\vex)-\delta]\rra }{\lla \delta_{Drc}[\delta(\vex)-\delta]\rra},
\eeq
where $\delta_{Drc}(\cdot)$ is Dirac's delta function and brackets
$\lla\cdot\rra$ 
represent to take ensemble average.

We assume that the  initial (linear) density fluctuations are isotropic
random Gaussian. At the linear-order we have $\veV(\vex)\propto \nabla
\Delta^{-1}\delta(\vex)$ and $\lla \veV(\vex)\delta(\vex)\rra=0$ due to
isotropy of matter fluctuations.  This means that
 $\delta$ and $\veV$ at a same point are statistically 
independent quantities, as a multivariate probability distribution function
(hereafter PDF) of Gaussian variables is completely decided by their
 covariance
matrix  ({\it e.g.} Bardeen et al. 1986).   Thus the
constrained velocity dispersion  
$ \Sigma_V^2 (\delta)$  does not
depend on  the density contrast 
$\delta$ at linear order.  However, nonlinear mode couplings would
 change the situation. Let us 
examine weakly non-Gaussian effects on $ \Sigma_V^2 (\delta)$. We can
express the first nonlinear correction of  $\Sigma_V^2 (\delta)$,  using
 framework of the
 Edgeworth expansion  method (Cramer 1946, Matsubara 1994, 1995,
 Juszkiewicz et al. 1995, Bernardeau \& Kofman 1995). This method is an
 excellent tool to explore 
 weakly nonlinear effects of the  large-scale structure induced by gravity. 

When  a field $F$ is defined by weakly non-Gaussian variables
$\{A_\mu(\vex)\}$ with vanishing means, we can expand the expectation
value $\lla F \rra$  as (see appendix A) 
\beq
\lla F(A_1,\cdots,A_m)\rra=\lla F \rra_G+\frac16 \sum_{\mu,\nu,\lambda}
\lla A_\mu A_\nu A_\lambda \rra_c \lla \frac{\p^3 F}{\p A_\mu\p A_\nu\p
  A_\lambda } \rra_G +O(\sigma^2 F),
\eeq
where $\lla \cdot \rra_G$ is the expectation value under the assumption that
variables 
$\{A_\mu(\vex)\}$ are multivariate Gaussian distributed, characterized by
their covariance matrix $\lla A_\mu A_\nu\rra$. The quantity
$\lla A_\mu A_\nu A_\lambda \rra_c$
is  the third-order  connected  
moment of  variables $\{A_\mu\}$,  and we have $\lla A_\mu A_\nu A_\lambda
\rra_c=\lla A_\mu A_\nu A_\lambda\rra$ at third-order.  The variance
$\sigma^2=O(A_i^2)$ is the order
parameter of perturbative expansion around  the Gaussian distribution, and 
we can regard  $\sigma^2=\lla 
\delta^2\rra$ in this article. The denominator of $\Sigma_V^2 (\delta)$
in equation (3) is nothing but the one point PDF of density contrast $\delta$.
From equation (4) we obtain the famous perturbative 
 formula as follows ($\nu\equiv \delta/\sigma$) 
\beq
\lla
\delta_{Drc}[\delta(\vex)-\delta]\rra=
\frac{e^{-\nu^2/2}}{{\sqrt{2\pi\sigma^2}}} 
  \lmk 1+\frac{S\sigma H_3(\nu )}{6} +O(\sigma^2)\rmk,
\eeq
({\it e.g.} Juszkiewicz et al. 1995, Bernardeau \& Kofman 1995) and this 
is the most
simplified version of the
Edgeworth expansion.
Here the function
$H_n(\nu)\equiv (-1)^ne^{\nu^2/2}(d/d\nu)^n e^{-\nu^2/2}$
is  $n$-th order Hermite polynomial, and $S$ is a parameter of order
unity and called skewness (Peebles 1980, Fry 1984, Goroff et al. 1986,
see also Seto 1999c), 
\beq
S\equiv \frac{\lla \delta^3\rra}{\sigma^4}.
\eeq
Due to the nonlinear correction term proportional to $S\sigma$, points with
high-$\sigma$ overdensity are more abundant than the linear prediction
by a  Gaussian
distribution.

Next the numerator of $\Sigma_V^2 (\delta)$ is given by
\beq
\lla\veV(\vex)^2
\delta_{Drc}[ \delta(\vex)-\delta]\rra=\sigma_V^2 
\frac{e^{-\nu^2/2}}{{\sqrt{2\pi\sigma^2}}} 
  \lmk 1+\frac{S\sigma H_3(\nu )}{6}+C\sigma
H_1(\nu) +O(\sigma^2)\rmk, 
\eeq
where $\sigma_V^2\equiv \lla\veV^2\rra$ is the unconstrained velocity
dispersion. The parameter $C=O(1)$ is defined by 
\beq
C=\frac{\lla \veV(\vex)^2\delta(\vex) \rra}
{\sigma^2 \sigma_V^2}.
\eeq
In the studies of the  large-scale structure, the Edgeworth expansion or 
the
third-order moments  have been mainly discussed for scalar fields, such as 
density field $\delta(\vex)$ or velocity divergence field $\nabla\cdot
\veV(\vex)$ ({\it e.g.} Chodorowski \& {\L}okas 1997). Here we present
analytical study for couplings  of 
$\delta(\vex)$ and $\veV(\vex)$, but numerical investigation of  our
method is also important as well as interesting.   
From equations (5) and (7) we obtain the constrained velocity dispersion 
$\Sigma_V^2 (\delta)$ up to the
first-order nonlinear correction as 
\beqa \displaystyle
\Sigma_V^2(\delta) &=& \frac{\lla \veV(\vex)^2\delta_{Drc}[\delta(\vex)-\delta]\rra }{\lla \delta_{Drc}[\delta(\vex)-\delta]\rra}\nn\\
&=& \frac{ \displaystyle\sigma_V^2 
\frac{e^{-\nu^2/2}}{{\sqrt{2\pi\sigma^2}}} 
  \lmk 1+\frac{S\sigma H_3(\nu )}{6}+C\sigma
H_1(\nu) +O(\sigma^2)\rmk}{ \displaystyle\frac{e^{-\nu^2/2}}{{\sqrt{2\pi\sigma^2}}} 
  \lmk 1+\frac{S\sigma H_3(\nu )}{6} +O(\sigma^2)\rmk}\nn\\
&=&\sigma_V^2\lmk1+\frac{S\sigma H_3(\nu )}6+C\delta-\frac{S\sigma H_3(\nu
    )}6+O(\sigma^2)\rmk \nn\\
&=&\sigma_V^2(1+C\delta+O(\sigma^2)).
\eeqa
Note that our result $\Sigma_V^2(\delta)$ does not depend on the skewness
parameter $S$. Nonlinear effects  appear through the quantity $C$.

Next let us evaluate non-Gausssianity induced by gravity,  using
higher-order perturbation theory.  We perturbatively
 expand the density and velocity fields as
\beqa
\delta(\vex)&=&\delta_1(\vex)+\delta_2(\vex)+\delta_3(\vex)+\cdots,\\
\veV(\vex)&=&\veV_1(\vex)+\veV_2(\vex)+\veV_3(\vex)+\cdots,
\eeqa
where $\delta_1(\vex)$ and $\veV_1(\vex)$ are  the linear modes,
$\delta_2(\vex)$ and $\veV_2(\vex)$ are the second-order modes, and so
on.   We solve the following three basic equations (continuity,
Euler and Poisson equations) order by order (Peebles 1980)
\beqa
\frac{\p}{\p
  t}\delta(\vex)+\frac{1}{a}\nabla[\veV(\vex)\{1+\delta(\vex)\}]&=&0,\\ 
\frac{\p}{\p
  t}\veV(\vex)+\frac1a[\veV(\vex)\cdot\nabla]\veV(\vex)+\frac{\p_t
  a}a\veV(\vex)+\frac1a\nabla\phi(\vex)&=&0,\\
\nabla^2\phi(\vex)-4\pi a^2\rho(t)\delta(\vex)&=&0,
\eeqa
where $a$ represents the scale factor. In these equations we have
omitted explicit time  dependence of fields for notational 
simplicities.  We only discuss quantities at a specific epoch and there
would be no confusion.

Fourier space representation is  convenient to analyze  the nonlinear
mode couplings. We denote the unsmoothed
 linear Fourier mode by $\delta_{lin}(\vek)$.
Then $\delta_1(\vex)$ and $\veV_1(\vex)$ are written in terms of
$\delta_{lin}(\vek)$ and $W(kR)$, the Fourier transform of the filter function
$W(|\vex|,R)$,  as 
 \beq
\delta_1(\vex)=\int \frac{d\vek}{(2\pi)^3}
\exp(i\vek\vex)\delta_{lin}(\vek)W(kR),~~~
\veV_1(\vex)={Hf}\int\frac{d\vek}{(2\pi)^3}  \frac{i\vek}{k^2} \exp(i\vek\vex)\delta_{lin}(\vek)W(kR),
\eeq
where $H(\equiv d\ln a/dt$) is the  Hubble parameter and 
 $f(\equiv d\ln D/d \ln a$,  $D$: linear growth rate of density
fluctuation) is a function of cosmological parameters $\Omega$ and
$\lambda$, and  well fitted by
\beq
f\simeq \Omega^{0.6}+\lambda/30,
\eeq
in the ranges  $0.05\le  \Omega \le 1.5 $ and $0 \le  \lambda \le 1.5 $
(Martel 1991).

We define the linear matter power spectrum $P(k)$ by
\beq
\lla \delta_{lin}(\vek)
\delta_{lin}(\vel)\rra=(2\pi)^3\delta_{Drc}^3(\vek+\vel)P(k).
\eeq
Then the dispersions
 $\sigma^2$ and $\sigma_V^2$ are given by the following simple integrals 
of $P(k)$ up to required order  to evaluate the first nonlinear effects of
 $\Sigma_V^2(\delta)$,
\beqa
\sigma^2&=&\int\frac{d\vek}{(2\pi)^3} P(k) W(kR)^2+O(\sigma^4),\\
\sigma_V^2&=&H^2f^2\int\frac{d\vek}{(2\pi)^3k^2} P(k) W(kR)^2+O(\sigma^4),
\eeqa
 In this 
article we only use the  Gaussian filter defined by  $W(kR)=\exp[-(kR)^2/2]$.

As shown in equation (8),  the first-order nonlinear correction of the
constrained dispersion 
$\Sigma_V^2(\delta)$ is 
characterized by the factor $C$. We need the second-order modes
$\delta_2(\vex)$ and $\veV_2(\vex)$ to calculate the first nonvanishing
contributions of 
$\lla\veV(\vex)^2\delta(\vex) \rra$.
These second-order modes  are given with linear mode
$\delta_{lin}(\vek)$ as (Fry 1984, Goroff 1986)
\beqa
\delta_2(\vex)&=&\int \frac{d\vek d\vel}{(2\pi)^6}
\exp[i(\vek+\vel)\vex]\delta_{lin}(\vek)
\delta_{lin}(\vel)\ch_{2\delta}(\vek,\vel) W(R|\vek+\vel|),\\ 
\veV_2(\vex)&=&{Hf}\int \frac{d\vek d\vel}{(2\pi)^6}
\frac{i(\vek+\vel)}{|\vek+\vel|^2} 
\exp[i(\vek+\vel)\vex]\delta_{lin}(\vek)
\delta_{lin}(\vel)\ch_{2V}(\vek,\vel)  W(R|\vek+\vel|), 
\eeqa

where kernels $\ch_{2\delta}$ and $\ch_{2V}$ are defined as follows
\beqa
\ch_{2\delta}(\vek,\vel)&=&\frac12(1+K)
+\frac{\vek\cdot\vel}{2kl}\lmk\frac{k}{l}+\frac{l}k\rmk +\frac12(1-K) 
\lmk \frac{\vek\cdot\vel}{kl}\rmk^2,\\ 
\ch_{2V}(\vek,\vel)&=&L
+\frac{\vek\cdot\vel}{2kl}\lmk\frac{k}{l}+\frac{l}k\rmk +(1-L) 
\lmk \frac{\vek\cdot\vel}{kl}\rmk^2.
\eeqa
The factors $K$ and $L$ depend very weakly on cosmological parameters
$\Omega$ and $\lambda$, and are fitted as (Matsubara 1995)
\beqa
K(\Omega,\lambda)&\simeq&\frac37\Omega^{-1/30}-\frac{\lambda}{80}\lmk1
-\frac32\lambda\log_{10}\Omega\rmk,\\ 
L(\Omega,\lambda)&\simeq&  \frac37\Omega^{-11/200}-\frac{\lambda}{70}\lmk1
-\frac73\lambda\log_{10}\Omega\rmk,
\eeqa
in the ranges  $0.1\le\Omega\le 1$ and $0\le\lambda\le 1$. In the
followings we neglect these weak dependence and simply put
\beq
K=L=\frac37.
\eeq
Thus we can write down the third-order moment 
$\lla \veV\cdot\veV \delta\rra$ in the following
form 
\beqa
\lla \veV\cdot\veV \delta\rra&=&
\lla \veV_1\cdot\veV_1\delta_2\rra+2\lla
\veV_1\cdot\veV_2\delta_1\rra+O(\sigma^6)\\ 
&=&2H^2f^2\int\frac{d\vek d\vel}{(2\pi)^6}P(k)P(l)\lkk
-\frac{\vek\cdot\vel}{k^2l^2}
\ch_{2\delta}(\vek,\vel)+2\frac{\vek\cdot(\vek+\vel)}{k^2|\vek+\vel|^2}\ch_{2V}
(\vek,\vel)\rkk \nn \\
& &\times W(kR)W(lR)W(|\vek+\vel|R)+O(\sigma^6).
\eeqa
Due to the rotational symmetry around the origin, we can simplify the
six dimensional integral $d\vek d\vel$ to three dimensional integral
$dkdldu$. Here, $-1\le u \le 1$ is the cosine between two vectors $\vek$
and $\vel$ and 
given by $u=\vek\cdot\vel/kl$. Then we obtain the first nonvanishing
order of $ \lla \veV\cdot\veV\delta\rra$ as
\beqa
\lla \veV\cdot\veV
\delta\rra&=&2H^2f^2\int_{-1}^1du\int\frac{k^2l^2dkdl}{8\pi^4}
P(k)P(l)\exp[-k^2-l^2-klu] \nn\\
& &\times \Bigg[-\frac{u}{kl}\lnk \frac57+\frac{u}2\lmk\frac{k}{l}+\frac{l}k\rmk+\frac27
u^2 \rnk\nn\\
& &~~~ +2\frac{k+lu}{k(k^2+l^2+2klu)}\lnk \frac37+\frac{u}2\lmk\frac{k}{l}+\frac{l}k\rmk+\frac47
u^2 \rnk   \Bigg].
\eeqa
Note that the parameter $C$ does not depend on the normalization of power
spectrum (see eqs.[18][19] and [29]).
Furthermore, the factors  $Hf$ cancel out between $\sigma_V^2$ and $\lla
\veV\cdot\veV \delta\rra$ and cosmological parameters are irrelevant for
the factor 
$C$ in our treatment ($K=L=3/7$). Finally,  we comment that  even though
the  constrained dispersion 
 $\Sigma_V^2(\delta)$ changes by $\sigma_V^2C\delta$ from the 
unconstrained value $\sigma_V^2$, the 
shape of the one-point PDF of velocity field with a given $\delta$
 keeps  Gaussian distribution at the same order of nonlinearity.  We can
 easily  confirm  
 this by calculating the ratio
\beq
\frac{\lla \delta_{Drc}^3(\veV(\vex)-\veV)
  \delta_{Drc}(\delta(\vex)-\delta)\rra}{\lla
  \delta_{Drc}(\delta(\vex)-\delta)\rra}
=\frac1{(2\cdot3^{-1}\pi\Sigma_V^2(\delta))^{3/2}}\lkk
\exp\lmk-\frac{\veV^2}{2\cdot3^{-1}\Sigma_V^2(\delta) } \rmk  +O(\sigma^2)\rkk.
\eeq
The factor $3^{-1}$ in the right-hand side arises from the dimensionality of 
the velocity vector.
First-order correction is completely absorbed to  the velocity dispersion 
$\Sigma_V^2(\delta)$.

\section{Results}
In this section we numerically evaluate the   parameter $C$ for various
power spectra. 
We first examine  pure power-law spectra $P(k)$ given 
by $(n >-1)$
\beq
P(k)=Ak^n.
\eeq
In this case $C$ does not depend on the smoothing radius $R$, and we can 
simply put $R=1$. Then the dispersions $\sigma^2$  and $\sigma_V^2$ are given as
\beq
\sigma^2=\frac1{(2\pi)^2}\Gamma \lmk \frac{3+n}2\rmk,~~~
\sigma_V^2=\frac{(Hf)^2}{(2\pi)^2}\Gamma \lmk \frac{1+n}2\rmk,
\eeq
where $\Gamma(n)$ is the Gamma function.
As for the nonlinear  coupling  
$\lla\veV\cdot\veV\delta\rra=\lla\veV_1\cdot \veV_1 \delta_2 \rra+2\lla\veV_1\cdot \veV_2 \delta_1
\rra$, we can write down the first 
 contribution 
$\lla \veV_1\cdot\veV_1\delta_2\rra$ explicitly in terms of Hypergeometric
functions as in the case of skewness parameter $S$ (Matsubara 1994,
{\L}okas et al 1995).  However, using {\it mathematica} (Wolfram 1996), 
we confirm  that the second term  $2\lla \veV_1\cdot\veV_2\delta_1\rra$
cannot be expressed in a closed form and numerical integration is required.
These two terms diverge in the limit $n\to -1$ where velocity dispersion 
$\sigma_V^2$ also
diverges, but  the factor $C$ approaches $0$ in this limit.

In Fig.1 we plot  $C$ as a function of spectral index $n$ in the range
$-1<n<2$.
The correction  $C$ is a positive and increasing function of $n$. 
This means that the velocity dispersion of high density regions are larger
than that of low density regions.
We have  $C= 0.314$  at the scale-invariant spectrum $n=1$.

Next we examine $C$ for a more realistic power spectrum $P(k)$. We use CDM 
 transfer function  given in Bardeen, Bond, Kaiser \& Szalay (1986) 
 and assume that the  primordial spectral
 index is equal to 1. Then $P(k)$  can be written  as
\beq
P(k)=Ak\lkk\frac{\ln(1+2.34q)}{2.34q}
\rkk^2[1+3.89q+(16.1q)^2+(5.46q)^3+(6.71q)^4 ]^{-1/2}, 
\eeq
where $q\equiv k/[(\Gamma h){\rm Mpc^{-1}}]$.  $\Gamma$ is the  shape
 parameter of the CDM transfer function 
and recent observational analyses of galaxy clusterings
 support $\Gamma = 0.2\sim 0.3$ ({\it e.g.} Tadros et al. 1999, Dodelson \& Gazta$\tilde{\rm n}$aga 1999).
In Fig.2 we plot $C$ as a function of smoothing radius $R$ 
in units of  $[(\Gamma h)^{-1}{\rm Mpc}]$.
 For this model the factor $C$ depends weakly on the smoothing radius
$R$ and we have $C\sim 0.30$ at 
a  weakly nonlinear regime $R\sim 10h^{-1}\mpc$.
In the limit $R\to \infty$, $C$ converges to $0.314$ which is the  same
 value of 
 $C$ for the power-law model with  $n=1$ presented
in Fig.1.  This is reasonable as we have
\beq
\lim_{k\to 0}\frac{P(k)}{k}=const, 
\eeq
for CDM models analyzed here.

Our results obtained so far are the  velocity dispersion for points
constrained by the 
matter  density contrast $\delta$. One might have interest in the velocity
dispersion constrained by the galaxy density contrast $\delta_g$.
Here, let us assume deterministic but nonlinear biasing relation for the 
smoothed galaxy
distribution $\delta_g(\vex)$ and the matter distribution $\delta(\vex)$ as
\beq
\delta_g(\vex)=b_1\delta(\vex)+b_2(\delta(\vex)^2-\sigma^2)+O(\sigma^3),
\eeq
where  $b_1$ and $b_2$ are some constants ({\it e.g.} Fry \& Gazta$\tilde{\rm n}$aga 1993). In this case we can easily
show that the velocity dispersion $\Sigma_V(\delta_g)^2$ for points $\vex$
with $\delta_g(\vex)=\delta_g$ is given by
\beq
\Sigma_V^2(\delta_g)=\sigma_V^2(1+C\delta_g/b_1+O(\sigma^2)),
\eeq
where the factors $C$ and  $\sigma_V$ are same as those appeared in
$\Sigma_V^2(\delta)$ (eq.[9]). Thus $\Sigma_V^2(\delta_g)$ does not
depend on the nonlinear coefficient $b_2$. This is also apparent when we
write down  $\delta(\vex)$ using $ \delta_g(\vex)$ and then insert this
solution to equation (9). The factor proportional to $b_2$ is higher
effects than analyzed here.  Note that in
 equation (36), the linear bias parameter $b_1$
appears by itself not in the usual form $\beta\equiv \Omega^{0.6}/b_1$,
and the overdensity $\delta_g$ dependence becomes smaller for larger  $b_1$.

Kepner et al. (1997) numerically investigated the mean magnitude
$\lla|\veV(\vex)|\rra$  of smoothed 
bulk velocity  for points with given overdensity
$\delta$.
Following the fact commented in the last paragraph of section 2, we can
easily calculate this magnitude $ \mu_V(\delta)$ and obtain following
result (see appendix B)
\beq
 \mu_V(\delta)=\sqrt{\frac8{3\pi}\Sigma_V^2(\delta)(1+O(\sigma^2))}=\sqrt{\frac8{3\pi}}\sigma_V \lmk 1+\frac{C\delta}2+O(\sigma^2)\rmk.
\eeq
For a typical CDM model, our analytical result predicts that the magnitude
$\mu_V(\delta)$ is expected to change $\sim 0.15\delta$, according to
the 
local density contrast $\delta$.  If we constrain points using overdensity
of galaxies instead of that of the gravitating matter, the combination 
$C\delta$ is replaced by 
$C \delta_g/b_1$ in the above equation.

Numerical results of Kepner et al. (1997) were given for
 CDM and HDM models 
with $\sigma_8=0.67$ normalization. Here $\sigma_8$ is the linear rms
density fluctuation in  a sphere of $8\hmpc$ radius.
They calculated the smoothed density and velocity fields with smoothing
radius $R$ at  $0.77\hmpc\le R\le 4.88\hmpc$. Thus their results are
 quantities at nonlinear regimes. It is true that simple application of our
perturbative formula to their results
would not be valid. However, surprisingly enough,
the quantity $\mu_V(\delta)$ shows almost no $\delta$ dependence in the
range $0<\delta\lsim 30$. \footnote{It  seems that the function
 $\mu_V(\delta)$ in their figures shows  extremely  weak
 dependence of $\delta$ around $\delta\simeq 0 $.}

If $\Sigma_V^2(\delta)$ (and thus $\mu_V(\delta)$) shows no
$\delta$-dependence at nonlinear scale and our second-order analysis is
valid for  
weakly nonlinear regime, we confront an  interesting possibility. 
Namely, with parameterization of overdensity by normalized value $\nu\equiv
\delta/\sigma$, velocity dispersion does not depend on $\nu$ at linear
and nonlinear regime, but depends on it  at (intermediate) weakly
nonlinear regime. Furthermore, we should notice that the 
velocity dispersion of dark-matter
particles (without coarse graining) depends largely on $\delta$ (Kepner
et al. 1997, Narayanan et al. 1998).\footnote{Definitions of velocity
  dispersion in these two papers are not identical. }

To make clear understanding of these transitions, 
we need to  numerically investigate the constrained dispersion
 $\Sigma_V^2(\delta)$ in
detailed manner with various smoothing length $R$, from linear to
nonlinear scales. Performance of second-order perturbation theory for
the 
velocity vector is also worth studying.

\section{Summary}
It is commonly accepted that the large-scale structure observed today 
 is formed by
gravitational instability from small primordial density fluctuations
(Peebles 1980).
In this picture, the peculiar velocity  and the density contrast  are
fundamental quantities to characterize inhomogeneities in the universe.
Linear analysis  of  cosmological perturbation theory predicts that, as long as
initial fluctuations are random Gaussian distributed,  the
one-point PDF of the velocity field $\veV(\vex)$ is statistically 
independent of the local density contrast   $\delta(\vex)$. This is an
 important 
aspect of cosmological    perturbation.

However nonlinear gravitational evolution changes the situation. Due to 
nonlinear mode-couplings, the  peculiar velocity field 
is no longer statistically
independent of the  local density field. 
 Here we have investigated bulk velocity dispersion
 ($\Sigma_V^2(\delta)$) as a function 
of the local density contrast  and calculated its first nonlinear correction
using framework of second-order perturbation theory.  Our target has
been set at
weakly nonlinear regimes  where perturbative treatment must be
reasonable. At present,  survey depth of velocity field is 
highly limited and our constrained statistics
 might  not be directly useful for
observational cosmology. However, we believe that our theoretical study
is important to understand one interesting aspect of the cosmic velocity field
peculiar to its nonlinear evolution.

We have shown  that the first nonlinear correction of
$\Sigma_V^2(\delta)$ is proportional to the local 
density and strongly depends on the matter power spectra, but weakly on the
cosmological parameters $\Omega$ and $\lambda$. For typical CDM model
with primordially scale-invariant fluctuations, this first-order
correction is about 
$0.3\delta$.  If we use  overdensity of galaxies $\delta_g$, this
correction term becomes $0.3 \delta_g/b_1$ (the factor $b_1$ is defined in
eq.[35]). This  dependence might be used to constrain the linear bias
parameter $b_1$ itself (not in the usual form $\Omega^{0.6}/b_1$) in future
peculiar velocity surveys. 
We have also shown that the constrained  one-point PDF of velocity field
keeps the Gaussian shape up to  second-order of perturbation. First
nonlinear effects are 
completely absorbed to 
the velocity dispersion $\Sigma_V^2(\delta)$. 

Numerical results by Kepner et al. (1997) have been compared with our 
analytical results.
Their results  show almost no $\delta$-dependence, 
contrary to ours. 
However spatial scale of their analysis (strongly nonlinear regime) is
largely different from ours (weakly nonlinear regime).
In the forthcoming paper, detailed numerical analysis 
will be presented for various smoothing lengths $R$ and power
spectra (see also Seto 2000). Validity of 
second-order  analysis as well as  the Edgeworth expansion method 
 for velocity vector 
would be also investigated numerically.

 \acknowledgments
The author would like to thank J. Yokoyama  
for useful discussion  and the referee R. Juszkiewicz for helpful comments.
This work was partially supported by the Japanese Grant
in Aid for Science Research Fund of the Ministry of Education, Science,
Sports and Culture  No.\ 3161.

\newpage
\appendix

\section{Weakly Non-Gaussian Averages}
In this appendix, we derive expression (4) for weakly non-Gaussian
variables $\{A_\mu\}~(\mu=1,\cdots,n)$ with $\lla A_\mu\rra=0$ ({\it
  e.g.} Matsubara 1994).
The partition function $Z(J_\mu)$ for a multivariate probability distribution function 
$P(A_\mu)$ is defined as
\beq
Z(J_\mu)\equiv \int_{-\infty}^{\infty} d^n P(A_\mu)\exp\lmk i \sum J_\nu A_\nu \rmk. 
\eeq
According to the cumulant expansion theorem (Ma 1985), the function $\ln
Z(J_\mu)$ is a generating function  of connected moments $\lla A_{\mu_1}
\cdots A_{\mu_N}\rra_c$. Therefore, taking the inverse Fourier transform of
equation (A1), the probability distribution function $P(A_\mu)$ is
written as
\beq
P(A_\mu)=\exp \lmk \sum_{N=3}^\infty \frac{(-1)^N}{N!}
\sum_{\mu_1,\cdots,\mu_N}\lla A_{\mu_1}
\cdots A_{\mu_N}\rra_c \frac{\p^N}{\p A_{\mu_1}  A\cdots\p A_{\mu_N} }
\rmk  P_G(A_\mu),
\eeq 
where the function $ P_G(A_\mu)$ is the multivariate Gaussian
probability distribution function determined by a ($n\times n$)
correlation matrix $M_{\mu\nu}\equiv \lla 
A_\mu A_\nu\rra$ as
\beq
P_G(A_\mu)=\frac1{\sqrt{(2\pi)^n {\rm det}M}}  \exp\lmk
-\frac12 \sum_{\mu,\nu} A_\mu (M^{-1})_{\mu\nu} A_\nu  \rmk .
\eeq
If we have relations  $\lla A_{\mu_1}
\cdots A_{\mu_N}\rra_c=O(\sigma^{2N-2}) $ as predicted by higher-order
perturbation theory, equation (A2) is perturbatively expanded as
\beq
P(A_\mu)=P_G(A_\mu)-\frac16 \sum_{\mu,\nu,\lambda} \lla A_\mu A_\nu
A_\lambda \rra_c   \frac{\p^3}{\p A_\mu\p A_\nu\p  A_\lambda} P_G(A_\mu) 
+O(\delta^2).
\eeq
Evaluating the  ensemble average of a field $F(A_\mu)$ with this
perturbative expression and taking partial integrals,  we
obtain  expansion (4). 

One might consider that formula (3) with Dirac's delta function
is somewhat indirect. We can obtain same results using formula
for probability distribution function $P(\veV|\delta)=P(\veV,\delta)/P(\delta)$
and evaluating $P(\veV,\delta)$ and $P(\delta)$ with expression (A4). 
\section{Derivation of $\mu_V(\delta)$}
Here we derive expression (37). As in the case of the constrained
velocity dispersion $\Sigma_V^2(\delta)$ (eq.[3]),  the quantity
$\mu_V(\delta)$ is formally defined as
\beq
\mu_V(\delta)\equiv \frac{\lla|\veV(\vex)|\delta_{Drc}[\delta(\vex)-\delta]\rra}{\lla\delta_{Drc}[\delta(\vex)-\delta]\rra}. 
\eeq
The r.h.s.  of this equation is written  as
\beq
\mu_V(\delta)=\int_{-\infty}^{\infty} d\veV  \frac{\lla |\veV| \delta_{Drc}^3(\veV(\vex)-\veV)
  \delta_{Drc}(\delta(\vex)-\delta)\rra}{\lla
  \delta_{Drc}(\delta(\vex)-\delta)\rra}.
\eeq
With the perturbative expansion given in equation (30) the r.h.s. of this equation is evaluated as
\beq
 \frac1{(2\cdot3^{-1}\pi\Sigma_V^2(\delta))^{3/2}}
\int_{-\infty}^{\infty} d\veV \lkk
\exp\lmk-\frac{\veV^2}{2\cdot3^{-1}\Sigma_V^2(\delta)  }  \rmk
+O(\sigma^2)\rkk |\veV|.
\eeq
It is straightforward to obtain  expression (37) given in the main
text as follows
\beq
 \mu_V(\delta)=\sqrt{\frac8{3\pi}\Sigma_V^2(\delta)(1+O(\sigma^2))}=\sqrt{\frac8{3\pi}}\sigma_V
 \lmk 1+\frac{C\delta}2+O(\sigma^2)\rmk.
\eeq

\newpage


\newpage
\begin{figure}[h]
 \begin{center}
 \epsfxsize=7.cm
 \begin{minipage}{\epsfxsize} \epsffile{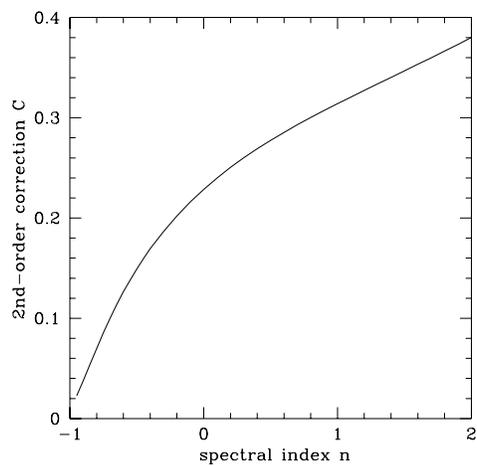} \end{minipage}
 \end{center}
\caption[]{The second-order correction $C$ for power-law matter
  spectra with Gaussian smoothing. }
\end{figure}

\begin{figure}[h]
 \begin{center}
 \epsfxsize=7cm
 \begin{minipage}{\epsfxsize} \epsffile{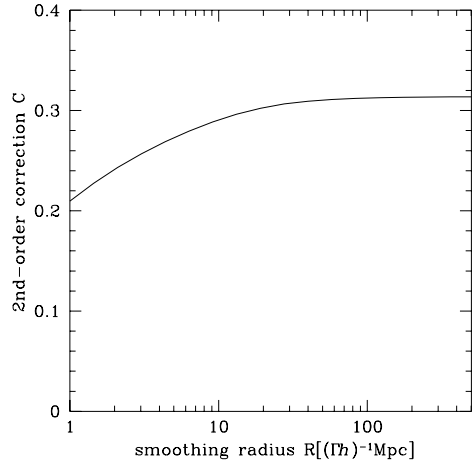} \end{minipage}
 \end{center}
\caption[]{The second-order correction $C$ for the CDM 
  spectrum of Bardeen et al. (1986). We plot the factor $C$ as a
  function of  the smoothing radius $R$ in
  units of  
  $(h\Gamma)^{-1}$Mpc. }
\end{figure}
\end{document}